\documentstyle[eqsecnum,prd,aps,floats,epsfig]{revtex}

\newcommand{\beq}{\begin{equation}}
\newcommand{\eeq}{\end{equation}}
\newcommand{\bea}{\begin{eqnarray}}
\newcommand{\eea}{\end{eqnarray}}

\def\laq{\raise 0.4ex\hbox{$<$}\kern -0.8em\lower 0.62
ex\hbox{$\sim$}}
\def\gaq{\raise 0.4ex\hbox{$>$}\kern -0.7em\lower 0.62
ex\hbox{$\sim$}}

\def \ra {\rightarrow}

\def \Da {\Delta}
\def \b {\beta}
\def \a {\alpha}
\def \ap {\alpha^{\prime}}
\def \Ga {\Gamma}
\def \ga {\gamma}
\def \sg {\sigma}
\def \da {\delta}
\def \ep {\epsilon}
\def \r {\rho}
\def \om {\omega}
\def \Om {\Omega}
\def \noi {\noindent}

\begin{document}

\def\baselinestretch{1.4} 

{\large
\begin{flushright}
CERN-TH/98-180\\
hep-ph/9806327
\end{flushright}
}

\vspace*{0.3truein}

\vskip 1.5 cm
{\Large\bf\centering\ignorespaces
Constraints on pre-big bang models for seeding large-scale \\
anisotropy  by massive Kalb--Ramond axions 
\vskip2.5pt}
{\dimen0=-\prevdepth \advance\dimen0 by23pt
\nointerlineskip \rm\centering
\vrule height\dimen0 width0pt\relax\ignorespaces

\vskip 1 cm
{\large  M. Gasperini}
\par}
{\large\it\centering\ignorespaces
Dipartimento di Fisica Teorica, Universit\`a di Torino,
Via P. Giuria 1, 10125 Turin, Italy \\
and Istituto Nazionale di Fisica Nucleare, Sezione di Torino,
Turin, Italy \\
\par}
{\dimen0=-\prevdepth \advance\dimen0 by23pt
\nointerlineskip \rm\centering
\vrule height\dimen0 width0pt\relax\ignorespaces
{\large G. Veneziano}
\par}
{\large\it\centering\ignorespaces
Theory Division, CERN, CH-1211 Geneva 23, Switzerland \\
\par}

\par
\bgroup
\leftskip=0.10753\textwidth \rightskip\leftskip
\dimen0=-\prevdepth \advance\dimen0 by17.5pt \nointerlineskip
\small\vrule width 0pt height\dimen0 \relax
 
\vskip 1.5 cm
\centerline{\Large Abstract}
\vskip 0.5 cm
\noi
{\large 
We discuss the conditions under which  pre-big bang  models can
fit the observed large-scale  anisotropy with a primordial
 spectrum of massive  (Kalb--Ramond) axion fluctuations.
 The primordial spectrum must be sufficiently flat at
low frequency and sufficiently steeper  at 
high frequency. For a steep and/or long enough 
high-frequency branch of the spectrum  the
bounds  imposed by  COBE's normalization allow axion masses of the
typical order for a Peccei--Quinn--Weinberg--Wilczek axion. We
provide a particular example in which an appropriate axion
spectrum is obtained from  a class of backgrounds satisfying the
low-energy string cosmology equations. 
}

\par\egroup

\vfill

{\large
\begin{flushleft}
CERN-TH/98-180\\
May 1998
\end{flushleft}
}


\def\baselinestretch{1}
\newpage
\setcounter{page}{1}

\par
\begingroup
\twocolumn[%

\begin{flushright}
CERN-TH/98-180\\
hep-ph/9806327\\
\end{flushright}
\vskip 0.5 true cm

{\large\bf\centering\ignorespaces
Constraints on pre-big bang models for seeding large-scale
anisotropy \\
by massive Kalb--Ramond axions 
\vskip2.5pt}
{\dimen0=-\prevdepth \advance\dimen0 by23pt
\nointerlineskip \rm\centering
\vrule height\dimen0 width0pt\relax\ignorespaces
 M. Gasperini
\par}
{\small\it\centering\ignorespaces
Dipartimento di Fisica Teorica, Universit\`a di Torino,
Via P. Giuria 1, 10125 Turin, Italy \\
and Istituto Nazionale di Fisica Nucleare, Sezione di Torino,
Turin, Italy \\
\par}
{\dimen0=-\prevdepth \advance\dimen0 by23pt
\nointerlineskip \rm\centering
\vrule height\dimen0 width0pt\relax\ignorespaces
G. Veneziano
\par}
{\small\it\centering\ignorespaces
Theory Division, CERN, CH-1211 Geneva 23, Switzerland \\
\par}
{\small\rm\centering(\ignorespaces May 1998\unskip)\par}

\par
\bgroup
\leftskip=0.10753\textwidth \rightskip\leftskip
\dimen0=-\prevdepth \advance\dimen0 by17.5pt \nointerlineskip
\small\vrule width 0pt height\dimen0 \relax

We discuss the conditions under which  pre-big bang  models can
fit the observed large-scale  anisotropy with a primordial
 spectrum of massive  (Kalb--Ramond) axion fluctuations.
 The primordial spectrum must be sufficiently flat at
low frequency and sufficiently steeper  at 
high frequency. For a steep and/or long enough 
high-frequency branch of the spectrum  the
bounds  imposed by  COBE's normalization allow axion masses of the
typical order for a Peccei--Quinn--Weinberg--Wilczek axion. We
provide a particular example in which an appropriate axion
spectrum is obtained from  a class of backgrounds satisfying the
low-energy string cosmology equations. 

\par\egroup
\vskip2pc]
\thispagestyle{plain}
\endgroup


\section{INTRODUCTION}
\label{I}

It has recently been shown \cite{1,1a} that a stochastic
background of massless axions can induce large-scale  anisotropies
of the Cosmic Microwave Background (CMB), in  agreement with
present observations \cite{2,2a}, provided it is primordially
 produced with a sufficiently flat spectrum. It has also been
shown that a massive axion   background can 
satisfy the same experimental constraints, with some
additional restrictions on the tilt of the spectrum, but only if the
axion mass lies inside an appropriate ultra-light mass window
\cite{1} having an upper limit of
 $10^{-17}$ eV.

The cosmic axion background considered in \cite{1,1a} is obtained by
amplifying the vacuum fluctuations of the so-called universal axion  
of string theory \cite{3}, i.e. the (four-dimensional) dual of the 
Kalb--Ramond (KR) antisymmetric tensor field  appearing in the
low-energy string  effective action. The KR axion has interactions
of gravitational strength, hence its mass is not significantly
constrained by  present tests of the equivalence principle for
polarized macroscopic bodies \cite{4}. Although (gravitationally) 
coupled to the QCD topological current, the KR axion 
 is not to be necessarily identified  with
the ``invisible" axion \cite{5} responsible for solving
 the strong CP problem. Other, more strongly coupled
 pseudoscalars, can play the traditional axion's role. In this
case, the standard Weinberg--Wilczek formula  \cite{6} would give
the mass of the appropriate combination of pseudoscalars which is
coupled to the topological charge, while the KR axion would 
mostly lie along the orthogonal combinations,  which
 remain (almost) massless. 

In this context it is thus  possible that
  KR axions are neither produced  from an initial
misalignment of the QCD vacuum angle \cite{7}, nor from the
decay of axionic cosmic strings \cite{8}, so that  existing
cosmological bounds on the axion mass \cite{9} can be evaded.
Also, KR axions are treated in \cite{1,1a} as ``seeds", i.e. as
inhomogeneous perturbations of a background that is {\it not}
axion-dominated, so that the mechanism of anisotropy production
is different from previous computations of isocurvature axion
perturbations \cite{10}. The seed approximation is not inconsistent 
either with the presence of mass or with the possible resonant 
amplification of quantum fluctuations through oscillations in the full 
axion potential \cite{kolb}. 

In spite of all this, it is quite likely that the KR axion
 will be heavier than  $10^{-17}$ eV, in which
 case, taking the results of \cite{1} at face value, KR axions
would  be unable to seed the observed 
CMB anisotropies. The main purpose of 
this paper is to point out that this conclusion is not inescapable, 
provided one is willing to add more structure to the primordial
axion spectrum   through more complicated cosmological
backgrounds than the one considered in   \cite{1}.

It  turns out, in particular, that the bounds on the axion mass can
be relaxed, provided the axion spectrum, before non-relativistic
corrections, grows   monotonically with frequency, but with a
frequency-dependent slope. At low frequencies   the slope must be 
small enough to reproduce the approximate scale-invariant
spectrum found by COBE, while at high frequencies the spectrum
has to be  steeper. Since position and normalization of
 the end point of the spectrum are basically
fixed in string cosmology, the steeper and/or longer
the  high-frequency spectrum, the larger the
suppression of the amplitude at the  low-frequency scales
relevant for  COBE's observation. 
On the other hand,  the low-frequency amplitude is
proportional to a positive power of 
the axion mass, in the non-relativistic regime. This
is the reason why, for a  steep and/or long  enough high-frequency
branch of the spectrum,  it becomes  possible to relax the bounds of
\cite{1} on the axion mass while remaining compatible
 with COBE's data.

In the context of the pre-big bang scenario \cite{11} it is
known that relativistic axions with a nearly flat spectrum can be
produced in the transition from a dilaton-dominated,
higher-dimensional phase to the standard radiation-dominated
phase \cite{3}. A spectrum with an effective  slope that grows
with frequency can be easily obtained if the phase of accelerated 
pre-big bang evolution consists of (at least) two distinct  regimes.
In that case the allowed mass window can  be enlarged
to include a more conventional range of values. 
Conversely, no significant
relaxation of the bounds given in \cite{1} seems to be possible 
if the two
branches of the spectrum  are obtained through a period of
decelerated post-big bang evolution, preceding the radiation era.

The rest of the paper is organized as follows. In Sections \ref{II} 
and \ref{III}   we generalize
the results of \cite{1,1a} for massive axions to an arbitrary
background of the pre-big bang type. In particular,
we will derive an equation expressing the predicted
CMB anisotropy  in terms of the axion mass, of the  string coupling,
 and of the behaviour of the cosmological
 background {\em before} the
radiation era. In Section \ref{IV} we will discuss in detail the
example of   an axion spectrum 
consisting of just a low- and a high-frequency branch. For this  
case we will determine the
region in parameter space that gives consistency with COBE's data
without   violating other important constraints. We will show
that backgrounds of   this kind can emerge, for instance,  from the
low-energy string cosmology equations in the   presence of 
classical string sources. Section \ref{V} is finally devoted to our
concluding remarks.

\section{MASSIVE AXION SPECTRA IN THE PRE-BIG
BANG SCENARIO}
\label{II}

We start by considering the dimensionally-reduced effective action
for Kalb--Ramond axion  perturbations ($\sg$), to lowest order in
$\ap$ and in the string coupling parameter. For
a spatially flat background, in the string frame, we are led to the
four-dimensional action \cite{3}: 
\beq
S={1\over 2} \int d^3x d\eta \left[ a^2 e^{\phi} \left(\sg^{\prime
2} + \sg\nabla^2 \sg\right) \right].
\label{21}
\eeq
Here a prime denotes differentiation with respect to conformal
time $\eta$, $\phi$ is the dimensionally reduced dilaton that 
controls the effective four-dimensional gauge coupling
$g=e^{\phi/2}$, and $a$ is the scale factor of the external,
isotropic three-dimensional space, in the string frame.
Variation of the action (\ref{21}) leads to the canonical
perturbation equations, which can be written in terms of the pump
field $\xi$ and of the normal mode $\psi$ as: 
\beq
\psi'' + \left(k^2-{\xi''\over \xi}\right)\psi=0,
~~~ \psi=\sg \xi, ~~~\xi=ae^{\phi/2}
\label{22}
\eeq
(we have implicitly assumed a Fourier expansion of perturbations,
by setting $\nabla^2 \psi=-k^2\psi$).

For any given model of background evolution,
$a(\eta)$, $\phi(\eta)$, the amplified axion spectrum
can  be easily computed, starting with an initial vacuum
fluctuation spectrum and  applying the standard formalism of
cosmological perturbation theory \cite{19}. One has to solve the
perturbation equation (\ref{22}) in the various cosmological
phases,  and to match the solution at the
transition epochs. From the final axion
amplitude $\sg (\eta)$, $\eta \ra +\infty$, one then obtains the
so-called Bogoliubov coefficients which determine, in the free-field,
oscillating regime (i.e. well inside the horizon), the total number
distribution $n(\om)$ of the produced axions.

For the purpose of this paper it will be enough to consider the case,
appropriate to the pre-big bang scenario, in which the pump field
$\xi$ keeps growing during the whole  pre-big bang
epoch, i.e. from $\eta=-\infty$ up to the final time $\eta=\eta_r$,
when the background enters the standard,
radiation-dominated regime with frozen dilaton. 
For $\eta>\eta_r$ the  pump
field is still growing, as the dilaton stays constant and $\xi(\eta)$
coincides with the expanding external scale factor $a(\eta)$.

With this model of background, it is convenient to refer the
spectrum  to the maximal amplified frequency, $\om_r
=k_r/a\simeq H_r a_r/a$, where $H=a'/a^2$ is the Hubble
parameter,  and $\om=k/a$ denotes proper frequency.
The axion energy distribution per logarithmic
interval of frequency, in this background, can
thus be written (in units of critical energy density $\r_c =
3H^2/8\pi G$) as:
 \bea
\Om_\sg(\om, \eta)&= &{1\over \r_c}{d \r\over d\ln \om}=
{4G\over 3 \pi H^2}\om^4 n(\om)\nonumber \\
&=& g_r^2 \Om_\ga (\eta)
\left(\om \over \om_r\right)^4 n(\om).
\label{28}
\eea
We have denoted with $\Om_\ga(\eta)=(H_r/H)^2(a_r/a)^4$ the
time-dependent radiation energy density (in critical units), that
becomes dominant at $\eta=\eta_r$. Also, we have identified the
curvature scale at the inflation--radiation transition with the
string mass scale, $H_r \simeq M_s$, and we have denoted by
$g_r\equiv g(\eta_r)$ the final value of the string coupling
parameter, approaching the present value of the fundamental ratio
between string and Planck mass \cite{17}:
\beq
g_r =e^{\phi_r/2} \simeq M_s/M_p \sim 0.1 - 0.01 .
\label{29}
\eeq
Numerical coefficients of order $1$ have been absorbed into
$g_r^2$, also in view of the uncertainty with which we can identify
the transition scale and the string scale.

The number distribution $n(\om)$  is completely determined by the
background evolution. For a monotonically growing pump field, as 
in the case we are considering, $n$ can be estimated as
\beq
n(\om)\simeq {\xi^2_{re}(\om)\over \xi^2_{ex}(\om)}=
{\xi^2_{r}\over \xi^2_{ex}(\om)}{a^2_{re}(\om)\over a^2_{r}}.
\label{26}
\eeq
Here the label $r$ denotes, as before, the beginning of the  
radiation phase; 
the labels ``{\it ex}" and ``{\it re}" mean evaluation of the
fields at the times $|\eta|\simeq (a\om)^{-1}$ when a mode $\om$,
respectively, ``exits the horizon" during the pre-big bang epoch,
and ``re-enters the horizon" in the post-big bang epoch.

The above estimate of $n(\om)$ 
has been performed by truncating the solution
of the perturbation equation (\ref{22}), ouside the horizon, to the
frozen part of the axion field, i.e. $\sg (\eta)=$ const for
$|k\eta|\ll 1$. This is common practice in the context of the
standard inflationary scenario \cite{19}, where the non-frozen
part of the fluctuations quickly decays in time outside the horizon.
It can be shown, however, that the estimate (\ref{26}) is generally
valid, quite independently of the behaviour of $\sg$ outside the
horizon, provided the total energy density is correctly computed
by including in the Hamiltonian the contribution of the frozen
modes  of the fluctuation and of its conjugate momentum
\cite{19a}.

Let us now discuss how the above spectrum has to be modified
when axions become massive. We will first assume that, at the
beginning of the radiation era,
 the axion field  has already acquired a  mass, but the mass  is so
small that it does not affect, initially, the axion spectrum. As the
Universe expands, however, the proper momentum is red-shifted
with respect to the rest mass, and a given axion  mode $k$ tends to
become non-relativistic when $\om=k/a<m$. The  spectrum
(\ref{28}) is thus valid only at early enough times, when the mass
contribution is negligible.

In order to include the late-time, non-relativistic corrections, we
may consider separately two different regimes.  If a mode
becomes non-relativistic well inside the horizon, i.e. when $\om
>H$, then the number $n(\om)$ of the produced axions is fixed
after re-entry, when the mode is still relativistic, and the
effect of the mass in the non-relativistic regime is a
simple rescaling of the energy density: $\Om_\sg \ra (m/\om)
\Om_\sg$. If, on the contrary, a mode becomes non-relativistic
outside the horizon,  when $\om <H$, then the final energy
distribution turns out to be determined by the background
kinematics at exit time
(as expected because of the freezing of the
fluctuations and of their canonical momentum outside the
horizon), and the effective  number of
non-relativistic axions has to be adjusted is such a way that
$\Om_\sg$ has the same spectral distribution as in the absence of
mass.  
The form of the 
non-relativistic corrections, in both regimes, can  be rigorously
obtained by solving the axion perturbation equation exactly with
the mass term included already from the beginning in the radiation
era \cite{1}, and also by 
using the general phenomenon of perturbation freeze out 
\cite{ramy} described in \cite{19a}. 

The two regimes are separated by the limiting frequency
$\om_m$ of a mode that  becomes non-relativistic just at the time
it re-enters the horizon \cite{1}. For modes re-entering during the
radiation era,
\bea
\om_m(\eta)&=&\om_r\left(m\over H_r\right)^{1/2} =
 \left(m H_{eq}\right)^{1/2} \Omega_{\gamma} \nonumber\\
& =&\left(m H \right)^{1/2} \Omega_{\gamma}^{1/4},
\label{211}
\eea
where the label $eq$ denotes, as usual, the time of  
matter--radiation equilibrium.
 Given the non-relativistic spectrum for $\om_m<\om<m$,
continuity at $\om_m$ then fixes $\Om_\sg$ for
the low-frequency band $\om<\om_m$.
Such a mass contribution to the spectrum was already taken into 
account when discussing non-relativistic corrections to the 
energy density of relic dilatons \cite{dil}. 

In order to compute the induced large-scale anisotropy, we need
the axion spectrum evaluated in the matter-dominated era, i.e. 
after the time $\eta_{eq}$. Also, we will assume that 
$ m>H_{eq}\sim 10^{-27}$eV 
(see Sect. \ref{III}), so that non-relativistic corrections are already
effective for $\eta >\eta_{eq}$. In the non-relativistic regime 
$\Om_\sg$ evolves in time like the energy density of dust matter,
 $\Om_\sg\sim a^{-3}$, and then, in the matter-dominated era, it
remains  frozen at the value $\Om_\sg(\eta_{eq})$ reached at
the time of matter--radiation equilibrium. By adding the
non-relativistic corrections to the generic spectrum  (\ref{28}) we
thus  obtain, for $\eta>\eta_{eq}$:
\bea
\Om_\sg(\om)&=&g_r^2\Om_\ga
\left(\om\over\om_r\right)^{4} n(\om),
~~~~~~~~~~~~~~~~~m<\om<\om_r,
 \nonumber\\
&=&g_r^2{m\over H_r}\left(H_r\over H_{eq}\right)^{1/2}
\left(\om\over\om_r\right)^{3} n(\om) ,
~~\om_m<\om<m, \nonumber\\
&=&g_r^2\left(m\over
H_{eq}\right)^{1/2} \left(\om\over\om_r\right)^{4} n(\om) ,
~~~~~~~\om<\om_m,\nonumber\\
&&\label{212}
\eea
where $n(\om)$ is the same axion number as in eqs. (\ref{28}),
i.e. the one determined by the solution of the relativistic
perturbation equation in the radiation era, and expressed in  terms
of the background as in eq. (\ref{26}). Also, we have used
$(H_r/H_{eq})^2(a_r/a_{eq})^3= (H_r/H_{eq})^{1/2}$. A particular
example of the spectrum  (\ref{212}) was already considered in
\cite{1}.

The particular shape of the spectrum is now determined by the
pump field $\xi(\eta)$, and thus by the background describing the
phase of pre-big bang evolution. Note
that the end-point value of the spectrum, $\Om_\sg (\om_r)$,
where by definition $n(\om_r)\simeq 1$ (according to eq.
(\ref{26})), is completely fixed in terms of the final string coupling
parameter only, $\Om_\sg (\om_r)=g_r^2 \Om_\ga \sim
10^{-4}g_r^2$. At the opposite (low-frequency) end, instead,  the
amplitude of the spectrum crucially depends on the axion mass, and 
the effect of non-relativistic corrections  is to enhance
the spectral amplitude at low frequency, as shown in Fig. 1.

\begin{figure}[t]
\begin{center}
\mbox{\epsfig{file=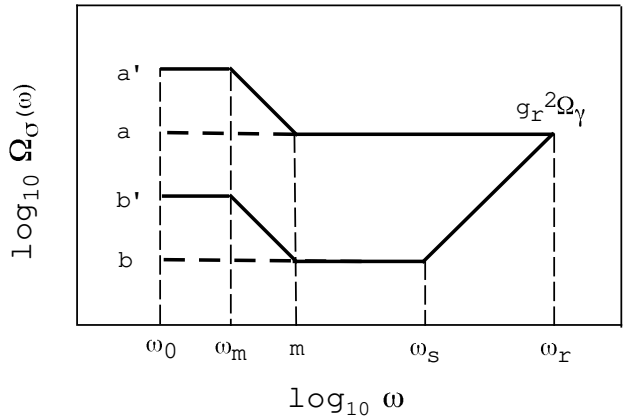,width=82mm}}
\vskip 5mm
\caption{\sl Two examples of axion spectra: ($a$) the limiting case
of a flat relativistic spectrum, with non-relativistic corrections
($a'$), and ($b$) the case of a relativistic spectrum which is flat
below $\om_s$, and grows above it, with non-relativistic
corrections ($b'$). The end-point at $\om_r$ is uniquely determined
by the string coupling parameter.}
\end{center}
\end{figure}

At fixed mass the low-frequency amplitude obviously depends on
the slope of the relativistic part of the spectrum, and thus on the
background.  In Fig. 1 we have plotted, for illustrative purposes, 
two possible spectra with non-relativistic corrections. The first
one,  labelled by {\sl (a)}, corresponds to a pre-big bang
background with $\xi \sim \eta^{-1}$, which leads to a flat
relativistic spectrum. The second one, labelled by {\sl (b)},
corresponds to a pre-big bang  background with $\xi \sim
\eta^{-1}$ up to $\eta_s$, and with  $\xi \sim \eta^{-1/2}$ for
$\eta_s<\eta<\eta_r$. The corresponding relativistic spectrum is
flat up to $\om_s \simeq (a\eta_s)^{-1}$, and grows linearly up to
$\om_r$. It is thus evident that the steeper is the slope of the
relativistic, high-frequency branch, the larger is the mass allowed
by the COBE normalization of the spectrum, as we will discuss in the
following sections.

The non-relativistic spectrum (\ref{212}) has been obtained 
under the
assumption that axions become massive at the beginning of the
radiation era \cite{1}. Actually, we may expect the effective axion
mass to turn on at some time during the radiation era, such as,
typically, at the deconfining/chiral phase transition. 
In that case, 
the above analysis remains valid provided axions become massive
before the frequency $\om_m$ re-enters the horizon. 

Let us call $T_m$ the temperature scale at which the mass turns on,
and $\om_T$ the proper frequency re-entering the horizon
precisely at the same epoch. The present value of $\om_T$ is then
\beq
\om_T(\eta_0) \simeq \om_{eq} \left(T_m\over {\rm eV}\right),
\label{omegat}
\eeq
 and the spectrum (\ref{212}) is
valid for $\om_m <\om_T$, namely for $m/H_{eq} <(T_m/{\rm
eV})^2$. 

In the opposite case, $\om_m >\om_T$, the role of the transition
frequency, that separates modes that become non-relativistic inside
and outside the horizon, is played by $\om_T$, 
and the lowest-frequency 
band of the spectrum (\ref{212}) has to be replaced by:
\bea
&&
\Om_\sg 
=g_r^2\left(m\over
H_{eq}\right) \left({\rm eV}\over T_m\right)
\left(\om\over\om_r\right)^{4} n(\om) , \nonumber\\
&&
\left(m\over
H_{eq}\right)^{1/2} \left({\rm eV}\over T_m\right)>1, 
~~~~~~~\om<\om_T
\label{temp}
\eea
(we have used  $\om_0\sim 10^{-2}\om_{eq} 
\sim 10^{-18}$ Hz, and $\om_r(t_0) \sim
g_r^{1/2} 10^{11}$ Hz). In this mass range the non-relativistic
spectrum is further enhanced with respect to eq. (\ref{212}) by the
factor $\left(m/
H_{eq}\right)^{1/2} \left({\rm eV}/ T_m\right)>1$, with a
consequently less efficient relaxation of the bounds on the axion
mass. This effect has to be taken into account when discussing
restrictions for a given model in parameter space. 

\section{Massive-axion contributions to $\Da T/ T$}
\label{III}

The main results of this section have been 
obtained also in \cite{1}, in the context of the ``seed" approach to
density fluctuations, by exploiting the so-called ``compensation
mechanism" \cite{19b} to estimate the relative contribution of
seeds and sources to the total  scalar perturbation
potential.  Here we will show, for the sake of completeness and for
the reader's convenience, that when the axion mass is sufficiently
large the contribution to the temperature anisotropies can be
quickly estimated also within the standard cosmological 
perturbation formalism, with results that are  the same as
those provided by the compensation mechanism. We also generalize
the results of \cite{1} to the case of a generic background
and spectrum.

We will work under the assumption that axions can be
treated as seeds for scalar metric perturbations:
the inhomogeneous axion stress tensor,
$\tau_\mu^{\nu}$, will  thus represent the total source of
perturbations, without contributing, however, to the unperturbed
homogeneous equations determining the evolution of the
background. Also, we shall only consider modes that are relevant
for the large-scale anisotropy, i.e. modes that are still outside the
horizon at the time of decoupling of matter and radiation,
$\eta_{dec} \sim \eta_{eq}$. Assuming that such modes are
already fully non-relativistic,
\beq
\om < H_{eq} <m,
\label{31}
\eeq
the corresponding axion stress tensor turns out to be completely
mass-dominated \cite{1}, so that we can neglect the off-diagonal
components and set
\beq
\tau_{ \mu}^\nu= {\rm diag} \left(\r_\sg,
-p_\sg \da_i^j\right).
\label{31a}
\eeq

In that regime, it will be convenient to write down the scalar
perturbation equations in the longitudinal gauge since, for
perturbations represented by a  diagonal stress tensor, the
perturbed metric $\da g_{\mu\nu}$ can be parametrized, in this
gauge, in terms of a single scalar potential $\Psi$ \cite{19}:
\beq
\da g_{\mu\nu}={\rm diag}~ 2 a^2 \Psi \left( 1, \da_{ij}\right).
\label{32}
\eeq
The perturbation of the Einstein equations, in the
matter-dominated era, then leads to relate the Fourier components
of $\Psi$ and of $\tau_{0}^0$ as \cite{19}:
\beq
\r_k = -{\r\over 6}\left[(k\eta)^2 \Psi_k +6\eta \Psi_k'+12
\Psi_k \right],
\label{33}
\eeq
where $\r$ is the total unperturbed matter energy density,
i.e. the source of the background metric, while
$\r_k=\tau_0^0(k)$ represents the contribution of the axion
background. In the matter era $\r\simeq \r_c$ so
that, for modes well outside the horizon ($k\eta \ll 1$),
\beq
\Psi_k \simeq -{1\over 3} {\r_k \over \r_c}.
\label{33a}
\eeq

In order to estimate the induced CMB anisotropies, we now
compute the so-called power spectrum of the Bardeen
potential, $P_\Psi (k)$, defined in
terms of the two-point correlation function of  $\Psi$ as:
\beq
\xi_\Psi (x,x')= \langle \Psi_x\Psi_{x'}\rangle=
\int {d^3k\over (2\pi k)^3}e^{i {\bf k} \cdot ({\bf x}-{\bf x}')}
P_\Psi(k)
\label{34}
\eeq
(the brackets denote spatial average or quantum expectation
values if perturbations are quantized).  The square root of
$\xi_\Psi$, evaluated at a comoving distance $|x-x'|=k^{-1}$,
represents the  typical  amplitude of  scalar metric
fluctuations induced by the axion seeds on a scale $k$.
 Using eq. (\ref{33a}),
\beq
P_\Psi^{1/2} (k) \simeq {1\over 3} {P_\r ^{1/2}(k)\over \r_c}\sim
{G\over H^2}P_\r^{1/2} (k),
\label{35}
\eeq
where $P_\r (k)$ is determined by the two-point correlation
function of the axion energy density, and then by the four-point
function of the stochastic, non-relativistic axion field $\sg(x)$:
\bea
&&
\int {d^3k\over (2\pi k)^3}e^{i {\bf k} \cdot ({\bf x}-{\bf x}')}
P_\r (k) =\nonumber\\
&&
=\langle \r_x(\sg)\r_{x'}(\sg)\rangle-
\langle\r_x(\sg)\rangle^2=
m^4\left(\langle \sg_x^2\sg_{x'}^2\rangle-
\langle\sg_x^2\rangle^2\right) \nonumber\\
&&
=m^4\int {d^3k\over (2\pi)^3}e^{i{\bf k}\cdot ({\bf x-x'})}
\int {d^3p\over (2\pi)^3}\left|\sg_p\right|^2
\left|\sg_{k-p}\right|^2.
\label{36}
\eea

The above integral is  extended, in principle, to all modes, both
the relativistic and the non-relativistic ones, in the three regimes
of  the axion spectrum. When considering the spectrum (\ref{212})
it turns out,
however, that for a sufficiently flat slope at low frequency the
integral over $p$ is dominated by the contribution of the region
$p\sim k$. Since we do need a flat spectrum (in order to fit the
observed large-scale anisotropy), and since we are restricting our
attention to all modes $k$ that re-enter after equilibrium,
$k<k_{eq}<k_m=k_{eq} (m/H_{eq})^{1/2}$, we can safely estimate
the integral through the contribution of the lowest frequency part
of the axion background $\sg_p$, i.e. for $p<k_m$. The same
conclusion applies to the lowest-frequency part of the spectrum
(\ref{temp}), since $k_{eq}<k_T\sim k_{eq}(T_m/{\rm eV})$. 
 We will now
compute the convolution (\ref{36}) for the particular case of the
spectrum (\ref{212}), but the result is also valid when the
axion mass is in the range corresponding to the spectrum
(\ref{temp}). 

We note, first of all, that in the frequency range we are
interested in, the effective 
axion field can be obtained from  eq. (\ref{212})
in the form
\bea
&&
\sg_p(\eta)\simeq {1\over a} \left[n(p)\over ma\right]^{1/2}
\left(p\over k_r\right)^{1/2}\left(H_r\over m\right)^{1/4},
\nonumber\\
&&
p<k_m, ~~~~~~~~~~~~~\eta>\eta_{eq},
\label{37a}
\eea
so that the integral (\ref{36}) becomes
\bea
&&
P_\r(k) \simeq \nonumber\\
&&
mH_r \left(k\over k_r\right)^{3}
\left( k_r\over a\right)^{6} \int {dp\over p} \left(p\over
k_r\right)^{4} {|p-k|\over k_r}n(p)n(|k-p|). \nonumber\\
&&
\label{37b}
\eea
We parametrize the slope of the relativistic axion spectrum by
 $p^4 n(p) \sim p^{3-2\mu}$,
with $\mu<3/2$ to avoid over-critical axion production. For a flat
enough slope, i.e. $\mu>3/4$, it can be easily checked that the
integrand of eq. (\ref{37b}) grows from $0$ to $k$, and decreases
from $k$ to $k_m$. The power spectrum may thus be immediately
estimated by taking the contribution at $p\sim k$, namely
\beq
P_\r(k) \sim
mH_r  \left( k_r\over a\right)^{6}\left(k\over k_r\right)^{8}
n^2(k).
\label{37c}
\eeq
Comparison with eq. ({\ref{212}) leads to the final result, valid for 
$\eta > \eta_{eq}$,
\bea
&&
P_\Psi^{1/2}(k) \sim {G\over H^2}P_\r^{1/2} (k)\nonumber\\
&&
\sim g_r^2 \left(m\over H_{eq}\right)^{1/2}
\left(k \over k_r\right)^{4} n(k)
\sim \Om_\sg(k),
\label{38a}
\eea
where $\Om_\sg$ is the lowest frequency, non-relativistic band of
the axion spectrum.

As the axion contribution to the
Bardeen potential  does not depend on time in the
matter-dominated era, the large-scale anisotropy
 of the CMB temperature is determined by the ordinary,
non-integrated  Sachs-Wolfe effect \cite{20} as
\cite{1,19}:
\beq
P_T^{1/2}(k)\sim P_\Psi^{1/2}(k) \sim \Om_\sg(k),
\label{38}
\eeq
where the temperature power spectrum $P_T(k)$ is defined by
\beq
\langle \Da T/T(x) \Da T/T(x')\rangle=
\int {d^3k\over (2\pi k)^3}e^{i {\bf k} \cdot ({\bf x}-{\bf x}')}
P_T (k).
\label{39a}
\eeq
Equation (\ref{38}) can  be converted (see \cite{1}) into  a relation for
the usual coefficients $C_{\ell}$ of the multipole expansion of
the CMB temperature fluctuations. 

Also, 
eqs. (\ref{38}), (\ref{38a}), and (\ref{26}) can be used to relate 
the cosmological background at a given time $\eta$ directly to the
axion mass and to the temperature anisotropy at a related scale.
We find: 
\beq
{\eta_{r}~\xi_{r}\over \eta ~\xi ({-\eta})} \simeq 
g_r^{-1}~\left(m\over H_{eq}\right)^{-1/4}
\left[P_T^{1/4}(k)\right]_{k\eta=1} , 
\label{relation}
\eeq
where we have used the fact that, for an accelerated power-law
background, the exit time of the mode $k$ in the pre-big bang
epoch is at $k \simeq -\eta^{-1}$. 

This equation can be seen as the analogue, in our context, of
the reconstruction  of the inflaton potential from the
CMB   power spectrum in ordinary slow-roll inflation  \cite{Kolb}.

 By parametrizing the slope of $\Om_\sg$ as $\Om_\sg \sim
\om^{(n-1)/2}$, it follows from eq. (\ref{38}) that $n$ can be
identified with the usual tilt parameter of the CMB anisotropy
\cite{2},   constrained by the data as:
\beq
0.8~\laq~n~\laq~1.4. 
\label{39}
\eeq
The  observed quadrupole amplitude, which normalizes the  spectrum
at the present horizon scale $\om_0$  \cite{2a}, gives also:
\beq
\Om_\sg(\om_0)\simeq 10^{-5}.
\label{310}
\eeq
In addition, the validity of our perturbative computation, which
neglects the back-reaction of the axionic seeds, requires that the
axion energy density remains well under-critical not only at the
end point $\om_r$ (which is automatically assured by
$g_r^2<1$), but also at the non-relativistic peak at $\om=\om_m$.
We thus require
\beq
\Om_\sg(\om_m)< 0.1.
\label{311}
\eeq

The COBE normalization (\ref{310}), imposed on the lowest
frequency end of the axion spectrum, fixes the mass as
a function of the background. In the two cases of eq. (\ref{212})
and (\ref{temp}) we get, respectively,
\bea
&&
\log_{10}{m\over H_{eq}} \simeq 
96- 2\log_{10}g_r-4 
\log_{10}\left[\xi_{r} \over \xi_{ex}(\om_0)\right], \nonumber\\
&&
~~~~~~\left(m\over H_{eq}\right)^{1/2} \left({\rm eV}\over T_m
\right) <1, \label{313}\\
&&
\log_{10}{m\over H_{eq}} \simeq 
48+\log_{10}\left(T_m\over {\rm eV} \right)- \log_{10}g_r
\nonumber\\
&&
-2\log_{10}\left[\xi_{r} \over \xi_{ex}(\om_0)\right], 
~~~~~~\left(m\over H_{eq}\right)^{1/2} \left({\rm eV}\over T_m
\right) >1. \label{313a}
\eea
In the next section it will be shown, with an
explicit example of background, that a very large axion mass
window may in principle be compatible with the bounds
(\ref{31}) and (\ref{39})--(\ref{311}).

\section{Constraints on parameter space for a particular class of
backgrounds}
\label{IV}

In order to provide a quantitative estimate of the possible axion
mass window allowed by the large-scale CMB anisotropy, in a
string cosmology context, we will discuss here a class of
backgrounds that is sufficiently representative for our purpose, and
characterized by three different cosmological phases. We parametrize  
the evolution of the
axionic pump field in these three phases  as
\bea &&
\xi \sim \left|\eta\right|^{r+1/2}, ~~~~~~~ \eta <\eta_s,
\nonumber\\
&&
\xi \sim \left|\eta\right|^{-\b}, ~~~~~~~~~~\eta_s<\eta <\eta_r,
\nonumber\\
&&
\xi \sim \left|\eta\right|, ~~~~~~~~~~~~~~ \eta_r <\eta ,
\label{25}
\eea
where $\eta_s$ marks the beginning of an intermediate phase,
preceding the standard radiation era, which starts at
$\eta=\eta_r$.  There is no need
to consider here also the last transition from 
radiation- to matter-dominance, 
at $\eta=\eta_{eq}$, since  we are
assuming $m>H_{eq}$, so that the axion spectrum becomes
mass-dominated (and thus insensitive to the subsequent
transitions) before the time of equilibrium. 

For the intermediate phase, $\eta_s<\eta<\eta_r$,  we have two
possibilities: accelerated or decelerated evolution of the
background fields, corresponding respectively to a shrinking or
expanding conformal time parameter, $|\eta_r/\eta_s|<0$ or
$|\eta_r/\eta_s|>0$. In the first case, as $\eta$ ranges from
$-\infty$ to $\eta_r$, the ``effective potential" $V=|\xi''/\xi|$
of eq. (\ref{22}) grows monotonically from zero to $V(\eta_r)\sim
\eta_r^{-2}$, the maximal amplified frequency is
$k_r=V^{1/2}(\eta_r)\sim \eta_r^{-1}$, and all frequency modes
``re-enter the horizon" in the radiation era.
In the second case $V$ is instead decreasing from $\eta_s$ to
$\eta_r$, the maximal amplified frequency is
$k_s=V^{1/2}(\eta_s)\sim \eta_s^{-1}$, and the high-frequency
band of the spectrum, $k_s<k<k_r$, re-enters the horizon during the
intermediate phase preceding the radiation era.

In the first case both the curvature scale and the string coupling 
$e^\phi$ keep growing. In the second case there is still a growth of
the coupling, but the curvature scale is decreasing. Such a phase
may correspond, typically, to the dual of the  dilaton-driven,
accelerated evolution from the string perturbative vacuum, and its
possible effects on the fluctuation spectra have already been 
discussed in \cite{15,16}.

In our case, as we shall see below, compatibility of
the low-frequency part of the spectrum with the
observed anisotropy requires the initial parameter $r$  to be
sufficiently near to $-3/2$ (this value simultaneously guarantees
a flat spectrum and an accelerated growth of the metric and of the
dilaton).  With this value of $r$, if we consider a
decelerated intermediate phase, it turns out that for $\b <-1$ and
$\b >2$ the high-frequency band of the axion spectrum is
decreasing, the amplification of perturbations is thus enhanced at
the COBE scale, and the bounds on the axion mass become more
constraining instead of being relaxed. For $-1 <\b<2$ the
spectrum grows monotonically, but an explicit computation shows
that even in that case no significant widening of the mass window
may be obtained.

In this paper we will thus concentrate on the first type of
background, namely on an accelerated evolution of the pumping
field,  parametrized as in eq. (\ref{25}). The axion spectrum grows
monotonically in the whole range  $-2<\b<1$, but it seems natural
to assume that also the pumping field is growing, so that we shall
analyse a background with $0<\b<1$. This background includes, in
the limit $\b \ra 1$, a phase of constant curvature and frozen
dilaton, $a \simeq (-\eta)^{-1}$, $\phi \simeq$ const, which is a
particular  realization of the high-curvature string phase introduced
in \cite{12,13} for phenomenological reasons, and shown to be a
possible late-time attractor of the cosmological equations when
the required higher-derivative corrections are added to the string
effective action \cite{16a}.

For the background (\ref{25}) the initial relativistic spectrum has
two branches, corresponding to modes that ``cross the horizon"
during the initial low-energy phase, $\om <\om_s \simeq H_s
a_s/a$, and during the subsequent intermediate phase,
$\om_s<\om<\om_r$. The non-relativistic corrections are to be
included according to eqs. (\ref{212}) and (\ref{temp}).
For the purpose of this paper, it will be sufficient to consider the
non-relativistic spectrum in two limiting cases only: the one
in which only the low-frequency branch of the spectrum
becomes non-relativistic, and the one in
which already in the high-frequency branch there are modes that
become non-relativistic outside the horizon. In the first 
case the spectrum is given by
\bea
\Om_\sg&=&g_r^2\Om_\ga\left(\om\over
\om_r\right)^{2-2\b},
~~~~~~~~~~~~~~~~~~~~~\om_s<\om<\om_r, \nonumber\\
&=&g_r^2\Om_\ga\left(\om\over
\om_r\right)^{3-2|r|}\left(\om_s\over \om_r\right)^{2|r|-2\b-1},
m<\om<\om_s, \nonumber\\
&=&g_r^2{m\over H_r}\left(H_r\over H_{eq}\right)^{1/2}
\left(\om\over\om_r\right)^{2-2|r|}
\left(\om_s\over \om_r\right)^{2|r|-2\b-1},  \nonumber\\
&& ~~~~~~~~~~~~~~~~~~~~~~~~~
~~~~~~~~~~~~~~~~~~~\om_m<\om<m,\nonumber\\
&=&g_r^2\left(m\over
H_{eq}\right)^{1/2}
\left(\om\over\om_r\right)^{3-2|r|}
\left(\om_s\over \om_r\right)^{2|r|-2\b-1}, \nonumber\\
&&~~~~~~~~~~~~~~~~~~~~~~~~~
~~~~~~~~~~~~~~~~~~~\om<\om_m,\label{213}
\eea
for $m/H_{eq}<(T_m/{\rm eV})^2$, and 
\bea
\Om_\sg&=&g_r^2\Om_\ga\left(\om\over
\om_r\right)^{2-2\b},
~~~~~~~~~~~~~~~~~~~~~\om_s<\om<\om_r, \nonumber\\
&=&g_r^2\Om_\ga\left(\om\over
\om_r\right)^{3-2|r|}\left(\om_s\over \om_r\right)^{2|r|-2\b-1},
m<\om<\om_s, \nonumber\\
&=&g_r^2{m\over H_r}\left(H_r\over H_{eq}\right)^{1/2}
\left(\om\over\om_r\right)^{2-2|r|}
\left(\om_s\over \om_r\right)^{2|r|-2\b-1},  \nonumber\\
&& ~~~~~~~~~~~~~~~~~~~~~~~~~
~~~~~~~~~~~~~~~~~~~\om_T<\om<m,\nonumber\\
&=&g_r^2\left(m\over
H_{eq}\right)\left({\rm eV}\over T_m\right)
\left(\om\over\om_r\right)^{3-2|r|}
\left(\om_s\over \om_r\right)^{2|r|-2\b-1}, \nonumber\\
&&~~~~~~~~~~~~~~~~~~~~~~~~~
~~~~~~~~~~~~~~~~~~~\om<\om_T,\label{213a}
\eea
for $m/H_{eq}>(T_m/{\rm eV})^2$. In the second limiting case the
spectrum is
\bea
\Om_\sg&=&g_r^2\Om_\ga\left(\om\over
\om_r\right)^{2-2\b},
~~~~~~~~~~~~~~~~~~~~m<\om<\om_r, \nonumber\\
&=&g_r^2{m\over H_r}\left(H_r\over H_{eq}\right)^{1/2}
\left(\om\over\om_r\right)^{1-2\b},
~~~~~~\om_m<\om<m, \nonumber\\
&=&g_r^2\left(m\over H_{eq}\right)^{1/2}
\left(\om\over\om_r\right)^{2-2\b},
~~~~~~~~~~~\om_s<\om<\om_m, \nonumber\\
&=&g_r^2\left(m\over H_{eq}\right)^{1/2}
\left(\om\over\om_r\right)^{3-2|r|}
\left(\om_s\over \om_r\right)^{2|r|-2\b-1}, \om<\om_s,
\nonumber\\
&& \label{214}
\eea
for $m/H_{eq}<(T_m/{\rm eV})^2$, and 
\bea
\Om_\sg&=&g_r^2\Om_\ga\left(\om\over
\om_r\right)^{2-2\b},
~~~~~~~~~~~~~~~~~~~~m<\om<\om_r, \nonumber\\
&=&g_r^2{m\over H_r}\left(H_r\over H_{eq}\right)^{1/2}
\left(\om\over\om_r\right)^{1-2\b},
~~~~~~\om_T<\om<m, \nonumber\\
&=&g_r^2\left(m\over H_{eq}\right)\left({\rm eV}\over T_m\right)
\left(\om\over\om_r\right)^{2-2\b},
~~~~~\om_s<\om<\om_T, \nonumber\\
&=&g_r^2\left(m\over
H_{eq}\right)\left({\rm eV}\over T_m\right)
\left(\om\over\om_r\right)^{3-2|r|}
\left(\om_s\over \om_r\right)^{2|r|-2\b-1}, \nonumber\\
&&~~~~~~~~~~~~~~~~~~~~~~~~~
~~~~~~~~~~~~~~~~~~~\om<\om_s,\label{214a}
\eea
for $m/H_{eq}>(T_m/{\rm eV})^2$.

The spectrum depends on five parameters: $g_r$,
$m$, $\om_s/\om_r$, $\b$ and $T_m$, the temperature scale at
which the axions become massive. The conditions to be imposed on
the axion energy density in order to fit present observations of the
large-scale anisotropy, without becoming over-critical, provide
strong constraints on the parameters $\om_s/\om_r$ and $\b$, 
namely on the pre-big bang evolution of the background, as we
now want to discuss. We will show that, for $\om_0<\om_s$, the
COBE normalization 
(\ref{310}) imposed on the lowest-frequency 
band of the above  spectra can be satified consistently
with all constraints  both for $(m/H_{eq})^{1/2}<T_m/$eV and 
$(m/H_{eq})^{1/2}>T_m/$eV, and the bounds on the mass can be
significantly relaxed.

In order to discuss this possibility, it is convenient to use as
parameters the duration of the intermediate pre-big bang phase,
measured by the ratio $\om_r/\om_s$, and the variation of the
pump field during that phase,
$\xi_s/\xi_r=(\eta_s/\eta_r)^{-\b}=(\om_r/\om_s)^{-\b}$. Thus
we set
\beq
\b=-{y\over x}, ~~x=\log_{10}(\om_r/\om_s)>0, ~~
y=\log_{10}(\xi_s/\xi_r)<0.
\label{41}
\eeq
According to eq. (\ref{39}), the slope of
the non-relativistic, low-energy branch of the spectrum,
\beq
(n-1)/2=3-2|r|,
\label{42}
\eeq
is constrained by
\beq
1.4 \leq |r|\leq 1.5
\label{43}
\eeq
 ($n<1$ has been excluded, to obtain a
growing axion spectrum also in the limit $\eta_s\ra \eta_r$). The
COBE normalization (\ref{310}) fixes the mass as follows
\bea
&&
\log_{10}{m\over
H_{eq}}=4y+\left(4|r|-2\right)x+164-116|r|\nonumber\\
&&
-\left(1+2|r|\right)\log_{10}g_r,
~~~~~~~~\left(m\over H_{eq}\right)^{1/2} \left({\rm eV}\over T_m
\right) <1, \label{44}\\
&&
\log_{10}{m\over
H_{eq}}=2y+\left(2|r|-1\right)x+\log_{10}\left(T_m\over {\rm
eV}\right) +82-58|r|\nonumber\\
&&
-\left({1\over 2}+|r|\right)\log_{10}g_r,
~~~~~~\left(m\over H_{eq}\right)^{1/2} \left({\rm eV}\over T_m
\right) >1. 
\label{44a}
\eea
The critical bound also depends on the mass: if 
$m/H_{eq}<(T_m/{\rm eV})^2$ the condition (\ref{311}) has to be
imposed on eq. (\ref{213}) for $\om_m<\om_s$, and 
 on eq. (\ref{214}) for $\om_m>\om_s$; if, on the contrary, 
$m/H_{eq}>(T_m/{\rm eV})^2$, then the condition (\ref{311}) has to
be imposed on eq. (\ref{213a}) for $\om_T<\om_s$, and 
 on eq. (\ref{214a}) for $\om_T>\om_s$. 
Finally, we have the constraints $\log_{10}
\left(m/H_{eq}\right)>0$, see eq. (\ref{31}), and, by definition,
$x>0$, $y<0$, $y>-x$ (since $\b <1$).

The allowed region in the  ($x,y$) plane, as determined by the above
inequalities, is not very sensitive to the variation of $g_r$ and $|r|$
in their narrow ranges, determined respectively by eqs. (\ref{29})
and (\ref{43}). For a qualitative illustration of the constraints
imposed by the COBE data we shall fix these parameters to the
typical values $g_r=10^{-2}$ and $|r|=1.45$ (corresponding to a
spectral slope $n=1.2$). Also, we will assume that axions become
massive at the scale of chiral symmetry breaking $T_m \simeq
100$ MeV. The corresponding allowed ranges of the
parameters of the intermediate pre-big bang phase (duration and
kinematics) are illustrated in Fig. 2.

\begin{figure}[t]
\begin{center}
\mbox{\epsfig{file=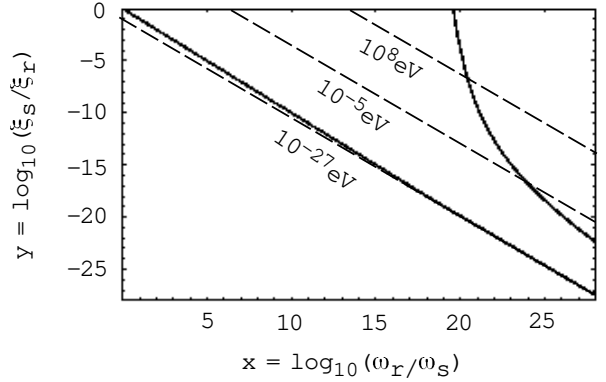,width=82mm}}
\vskip 5mm
\caption{\sl Possible allowed region for the parameters of an
intermediate  pre-big phase, consistent with an axion spectrum
that does not become over-critical, and that reproduces the present
COBE observations. The dashed lines represent curves of constant
axion mass.}
\end{center}
\end{figure}

The allowed region is bounded by the bold solid curves. The lower
border is fixed by the condition $m>H_{eq}\sim 10^{-27}$ eV for
$x~\gaq ~18$, and by the condition $\b<1$, i.e. $y>-x$, for
$x~\laq ~18$. The upper border is fixed by the condition
$\Om_\sg(\om_m)<0.1$, imposed for $\om_T>\om_s$ and
$m/H_{eq}>10^{16}$. 
The region is limited to the range
$\om_r/\om_s~\laq~10^{28}$, as we are considering the case in
which the axion contribution to the CMB anisotropy arises from the
low-frequency branch of the spectrum.

The dashed lines of Fig. 2 represent curves of constant axion mass,
determined by the conditions (\ref{44}), (\ref{44a}), with
$g_r=10^{-2}$, $|r|=1.45$ and $T_m=100$ MeV, namely
\bea
&&
y=-0.95x -0.9 +{1\over 4}\log_{10}\left(m\over 10^{-27} {\rm
eV}\right), ~m<10^{-11}~{\rm eV}, \nonumber\\
&&
y=-0.95x -4.9 +{1\over 2}\log_{10}\left(m\over 10^{-27} {\rm
eV}\right), ~m>10^{-11}~{\rm eV}.\nonumber\\
\label{48}
\eea
As shown in the picture, the allowed region is compatible with
masses much higher than $H_{eq}$, up to the limiting value $m\sim
100$ MeV above which the discussion of this paper cannot be
applied, since for $m>100$ MeV all the produced axions decayed
into photons before the present epoch, at a rate $\Ga \sim
m^3/M_p^2$. For the particular example shown in Fig. 2, the
allowed axion-mass window is then
\beq
10^{-27} ~{\rm eV} <m<10^2 ~{\rm MeV}.
\label{49}
\eeq
In the absence of the intermediate phase, i.e. for $x,y\ra 0$, we
recover the result obtained in \cite{1}, with $m\sim 10^{3.8}
H_{eq}\sim 10^{-23}$ eV.

A high-curvature string phase with nearly constant dilaton and
curvature scale, i.e. $\b \simeq 1$, $y\simeq -x$, is not excluded
but must lie very near the lower border of the allowed region, so
that it does not relax in a significant way the bounds on the axion
mass. It is nevertheless remarkable that such a phase is not
inconsistent with axion production and with the COBE normalization
of the axion spectrum. Such a phase would produce a relic gravity
wave background with  a nearly flat spectrum \cite{13}, easily
observable by advanced detectors, and its extension in frequency
would be constrained by $x~\laq~18$, to avoid conflicting with
present pulsar-timing data \cite{21}. This does not introduce
additional constraints in our discussion, however, since for $x>18$ the
line $y=-x$ lies outside the allowed region of Fig. 2.

We conclude this section by giving a possible example of background, 
which satisfies the low-energy string cosmology equations, and
which is simultaneously compatible with COBE and with higher
values of the axion mass, as illustrated in Fig. 2.

Consider the gravi-dilaton string effective action, with zero dilaton
potential, but with the contribution of additional matter fields
(strings, membranes, Ramond forms, ...) that can be approximated as
a perfect fluid with an appropriate equation of state.  As  discussed
in \cite{22}, the cosmological equations can in this case be 
integrated exactly, and the general solution is characterized by two
asymptotic regimes. In the initial small-curvature limit,
approaching the perturbative vacuum, the background is dominated
by the matter sources.  At late times, when approaching the
high-curvature limit, the background becomes instead
dilaton-dominated, and the effects of the matter sources
disappear. The time-scale marking the transition between the two
kinematical regimes (and thus the duration of the second,
dilaton-dominated phase) is controlled by an arbitrary integration
constant.

In order to provide an explicit example of this class of
backgrounds, we can take, for instance, a
$4+n$ manifold and we set
\bea
&&
g_{\mu\nu}= {\rm diag} \left(1, -a^2 \da_{ij}, -b^2 \da_{mn}\right),
\nonumber \\
&&
T_{\mu}^\nu= {\rm diag}~\r \left(1, -\ga \da_i^j,
-\ep \da_m^n\right),
\nonumber\\
&&
a\sim |t|^{\a_1}, ~~ b\sim |t|^{\a_2}, ~~
\phi =\Phi -n \ln b.
\label{410}
\eea
Here $\Phi$ is the unreduced, $(4+n)$-dimensional dilaton field and
we have called $a$ and $b$, respectively, the
external and internal scale factors, while $\ga$ and $\ep$ define the
external and internal equations of state. In the initial,
matter-dominated regime ($\eta<\eta_s$),
 the string cosmology equations lead to
\cite{22}:
\beq
r={5 \ga-1\over 1+3\ga^2+n\ep^2-2\ga}-{1\over 2}.
\label{411}
\eeq
In the subsequent dilaton-dominated regime ($\eta>\eta_s$) the
equations give \cite{23}
\beq
\b={1-3\a_1\over 1-\a_1}-{1\over 2}, ~~~~ \a_1^2
={1\over 3}\left(1-n\a_2^2\right),
 \label{412}
\eeq
where $\a_1, \a_2$ depend on $\ga$, $\ep$,
and on arbitrary integration constants.

The value $r=-3/2$, required for a flat
low-frequency branch of the spectrum, is thus obtained provided
internal and external pressures are related by:
\beq
n\ep^2 =-3\ga (1+\ga).
\label{413}
\eeq
On the other hand, an appropriate equation of state, motivated by
the self-consistency of this
background with the solutions of the string equations of motion
\cite{22,24}, suggests for $\ga$ the range $-1/3\leq \ga \leq 0$.
This range, together with the condition (\ref{413}),
also guarantees the validity of the so-called dominant energy
condition, $\rho \geq 0$, for the whole duration of the
low-energy pre-big bang phase. Near the singularity,
when the background enters the dilaton-dominated regime,
the kinematics, and then the value of $\b$, depends on the
integration constants. For a particularly simple choice of
such constants ($x_i=0$ in the notation of \cite{22}), one finds
$\a_1 = \sqrt{-\ga /3}$, and the value of $\b$ becomes
completely fixed by $\ga$ as
\beq
\b= { 1 + \sqrt{-3\ga} \over 1 + \sqrt{-\ga/3}}
-{1\over 2}.
\label{417}
\eeq
With $\ga$ ranging from $-1/3$ to $0$,
$\b$ ranges from $-1/2$ to
$1/2$. It is thus always possible, even in this simple
example, to implement the condition $\b<1$, in such a way as
to satisfy the properties required by the allowed region
of Fig. 2.

\section{Conclusion}
\label{V}

In this paper we have discussed, in the context of the pre-big bang
scenario, the possible consistency of a pseudoscalar origin of the
large-scale anisotropy, induced by the fluctuations of
non-relativistic Kalb--Ramond axions, with masses up to the $100$
MeV range. The enhancement of the low-energy tail of the axion
spectrum, due to their mass, has been shown to be possibly
balanced by the depletion induced by a steeper slope at high
frequency.  We have provided an explicit example of background
that satisfies the low-energy string cosmology equations, and leads
to an axion spectrum compatible with the above requirements.

The discussion of the reported example is incomplete in many
respects. For instance, the class of models that we have
considered could be generalized by the inclusion of additional
cosmological phases; also, an additional reheating subsequent to
the pre-big bang $\ra$ post-big bang transition could dilute the
produced axions, and relax the critical density bound; and so on.
In this sense, the  results discussed in Sect. \ref{IV} are to be taken
only as indicative of a possibility.

In this spirit, the main message of this paper is that in the context
of the pre-big bang scenario there is no fundamental physical
obstruction against an axion background that fits consistently the
anisotropy observed by COBE,  with ``realistic" masses in the
expected range of conventional axion models \cite{5} --\cite{9}.

For a given axion mass, the corresponding anisotropy is only a
function of the parameters of the pre-big bang models. If the
axion mass were independently determined, the measurements
of the CMB anisotropy might be interpreted, in this context, as
indirect observations of the properties of a very early cosmological
phase, and might provide useful information about the
high-curvature, strong-coupling regime of the string cosmology
scenario.

\acknowledgements

We are grateful to Ruth Durrer and Mairi Sakellariadou for helpful
discussions.

\end{document}